\documentclass[%
 reprint,
 amsmath,amssymb,
 aps,
]{revtex4-2}

\usepackage{graphicx}
\usepackage{dcolumn}
\usepackage{bm}
\usepackage{natbib}
\usepackage{multirow}
\usepackage{quantikz}
\usepackage{braket}
\usepackage{hyperref}
\usepackage{booktabs}
\usepackage{siunitx}
\usepackage{amsmath}
\usepackage{graphicx}

\usepackage{subcaption}

\usepackage{tikz}

\newcommand{\plum}{
\begin{tikzpicture}[scale=0.2]
\fill[purple] (0,0) circle (0.5);
\draw[green!60!black, thick] (0.2,0.5) -- (0.5,0.9);
\end{tikzpicture}
}

\definecolor{color1}{HTML}{648FFF}
\definecolor{color2}{HTML}{785EF0}
\definecolor{color3}{HTML}{DC267F}
\definecolor{color4}{HTML}{FE6100}
\definecolor{color5}{HTML}{FFB000}
\definecolor{color6}{HTML}{882255}
\definecolor{color7}{HTML}{0C7BDC}
\definecolor{white}{HTML}{FFFFFF}

\begin{document}

\preprint{APS/123-QED}

\title{Particle-Lund Multimodality in Jet Taggers}

\author{Loukas Gouskos}
\email{loukas.gouskos@cern.ch}
\affiliation{%
  Brown University, Providence, USA
}%

\author{Benedikt Maier}
\email{benedikt.maier@cern.ch}
\affiliation{%
 Imperial College of Science, Technology and Medicine, London, United Kingdom
}%

             
\begin{abstract}
The Lund plane offers a physics-motivated, hierarchical representation of QCD radiation within jets, while transformer-based taggers have reached state-of-the-art performance by learning directly from raw particle constituents and their pairwise relations. We investigate whether transformers implicitly capture hierarchical QCD structure from constituent-level inputs, or whether explicit physics representations remain complementary. To test this, we introduce PLuM\plum, a multimodal architecture that projects particle constituents and Lund plane splittings into a shared latent space, processing both jointly with a unified transformer. Cross-attention allows the model to probe whether structured QCD information provides discriminating power beyond what particles alone encode. We observe systematic gains for top-quark and $\mathrm{H}\to\mathrm{b}\bar{\mathrm{b}}$ tagging, while finding no comparable improvement for $\mathrm{H}\to\mathrm{c}\bar{\mathrm{c}}$ or $\mathrm{H}\to 4$q topologies. This selective enhancement suggests that explicit hierarchical information about b-jet formation remains complementary to raw particle representations even in highly expressive architectures, while other topologies are already well-captured at constituent level. For high-impact LHC analyses such as Lorentz-boosted di-Higgs searches in the four $\mathrm{b}$ quark final state ($\mathrm{H}\mathrm{H}(4\mathrm{b})$), the gains are substantial: at a 25\% di-Higgs efficiency working point, PLuM achieves 25\% higher background rejection than the baseline. Our results indicate that physically structured representations of QCD radiation retain discriminating value in the transformer era, motivating further study into how different aspects of jet dynamics are encoded by deep learning algorithms.

\end{abstract}

\maketitle


\section{\label{sec:level1}Introduction}

In recent years, the application of machine learning (ML) to jet physics at the Large Hadron Collider (LHC) has led to substantial improvements in classification and tagging performance. Among the most successful developments are graph-based and transformer-based models, such as ParticleNet~\cite{Qu:2019gqs} and Particle Transformer (ParT)~\cite{Qu:2022mxj}, which treat jets as sets of constituent particles and learn to capture complex correlations via message-passing mechanisms such as edge convolutions~\cite{DBLP:journals/corr/abs-1801-07829} and attention~\cite{DBLP:journals/corr/VaswaniSPUJGKP17}. Especially transformers are capable of learning representations directly from low-level, data-driven inputs without requiring handcrafted features or strong inductive biases. They have become the state of the art for jet tagging at ATLAS and CMS~\cite{,ATLAS:2025dkv,CMS:2026uph}, but also have applications far beyond jet analysis~\cite{Maier:2021ymx,Spinner:2024hjm,Quetant:2024ftg,Caron:2024cyo,Brehmer:2024yqw,VanStroud:2024fau,Koay:2025bmu}.

On the other hand, the Lund jet plane~\cite{Dreyer:2018nbf,Ghira:2025nym} provides a structured and interpretable representation of a jet’s internal dynamics, capturing the phase-space distribution of sequential emissions in a way that reflects the singular behavior of QCD. Originally introduced to aid in analytic understanding of jet substructure, the Lund plane has more recently been proposed as a physics-motivated feature space for ML-based applications~\cite{Dreyer:2020brq,Dreyer:2021hhr,Diaz:2023otq}. It  offers a complementary view to particle-level inputs and has been extensively probed by the LHC experiments in recent years~\cite{CMS:2023lpp,ATLAS:2024dua,LHCb:2025mcq,CMS:2025eyd,CMS:2025sch,CMS:2026bmz,CMS:2026yrb}.

In this work, we investigate whether the information encoded in the Lund jet plane provides complementary information beyond what is already learned by state-of-the-art transformer architectures. While particle-based transformers directly learn from low-level kinematic inputs and pairwise relations, the Lund plane introduces an explicit representation of the hierarchical QCD radiation pattern. The central question is therefore whether modern transformer architectures already saturate the information content of particle-level representations or whether physically structured views of the same jet provide additional discriminating power.

Specifically, we augment a ParT-style transformer with features derived from the Lund plane and study the impact on classification performance in two benchmark tasks: top quark vs. QCD and H vs. QCD jet tagging, where the Higgs boson decays into $\mathrm{b}\bar{\mathrm{b}}$, $\mathrm{c}\bar{\mathrm{c}}$, or $4\mathrm{q}$ (via $\mathrm{WW}^*$).

Our study explores the interplay between data-driven and physics-informed representations and examines whether combining both views within a common latent space can provide additional sensitivity in collider analyses. More generally, it probes whether explicitly encoding known structures of QCD radiation can complement highly expressive attention-based architectures.

\section{Lund Plane Features}

The Lund plane is a powerful representation of the internal structure of jets, inspired by the QCD branching process. It provides a two-dimensional map of jet splittings, constructed from the sequential clustering of jet constituents using algorithms such as Cambridge--Aachen (C/A)~\cite{ca1,ca2} or $k_\mathrm{T}$~\cite{kt}. Each branching in the clustering tree is mapped onto the Lund plane using the following kinematic variables:

\begin{itemize}
    \item \textbf{Angle} ($\Delta R$): the angular separation between the two branches of a splitting,
    \[
    \Delta R = \sqrt{(\Delta \eta)^2 + (\Delta \phi)^2},
    \]
    where $\eta$ and $\phi$ are the pseudorapidity and azimuthal angle.

    \item \textbf{Energy sharing} ($z$): the momentum sharing between the two subjets,
    \[
    z = \frac{\min(p_{\mathrm{T}1}, p_{\mathrm{T}2})}{p_{\mathrm{T}1} + p_{\mathrm{T}2}},
    \]
    where $p_{\mathrm{T}1}$ and $p_{\mathrm{T}2}$ are the transverse momenta of the branches.

    \item \textbf{Transverse momentum scale} ($k_\mathrm{T}$): the relative transverse momentum at the splitting,
    \[
    k_\mathrm{T} = z \cdot p_\mathrm{T} \cdot \Delta R,
    \]
    where $p_\mathrm{T}$ is the transverse momentum of the parent pseudojet before the splitting.
\end{itemize}

These splittings can be used to construct a graph-based representation of the jet, where nodes correspond to individual branchings and edges follow the clustering history. Each node can be described by the logarithms of its $(z, k_\mathrm{T}, \Delta R)$ tuple, providing a compact and physically-motivated description of the jet substructure.

Compared to particle-based graphs, where each node corresponds to a final-state particle, Lund plane representations provide several advantages for jet tagging:
\begin{itemize}
    \item They explicitly encode the hierarchy of QCD radiation through the clustering tree.
    \item The use of logarithmic coordinates $\log(1/\Delta R)$ and $\log(k_\mathrm{T})$ stretches the phase space in the soft/collinear region, where a lot of the discriminating power for quark/gluon separation or boosted object tagging lies.
    \item The representation is typically more compact, with fewer nodes than particle-based graphs, providing a highly dense representation that captures key physics features.
    \item The inputs reflect QCD priors directly, potentially improving interpretability.
\end{itemize}

\section{Adding the Lund plane}

\begin{figure*}[t!]
\begin{centering}
\includegraphics[width=0.975\textwidth]{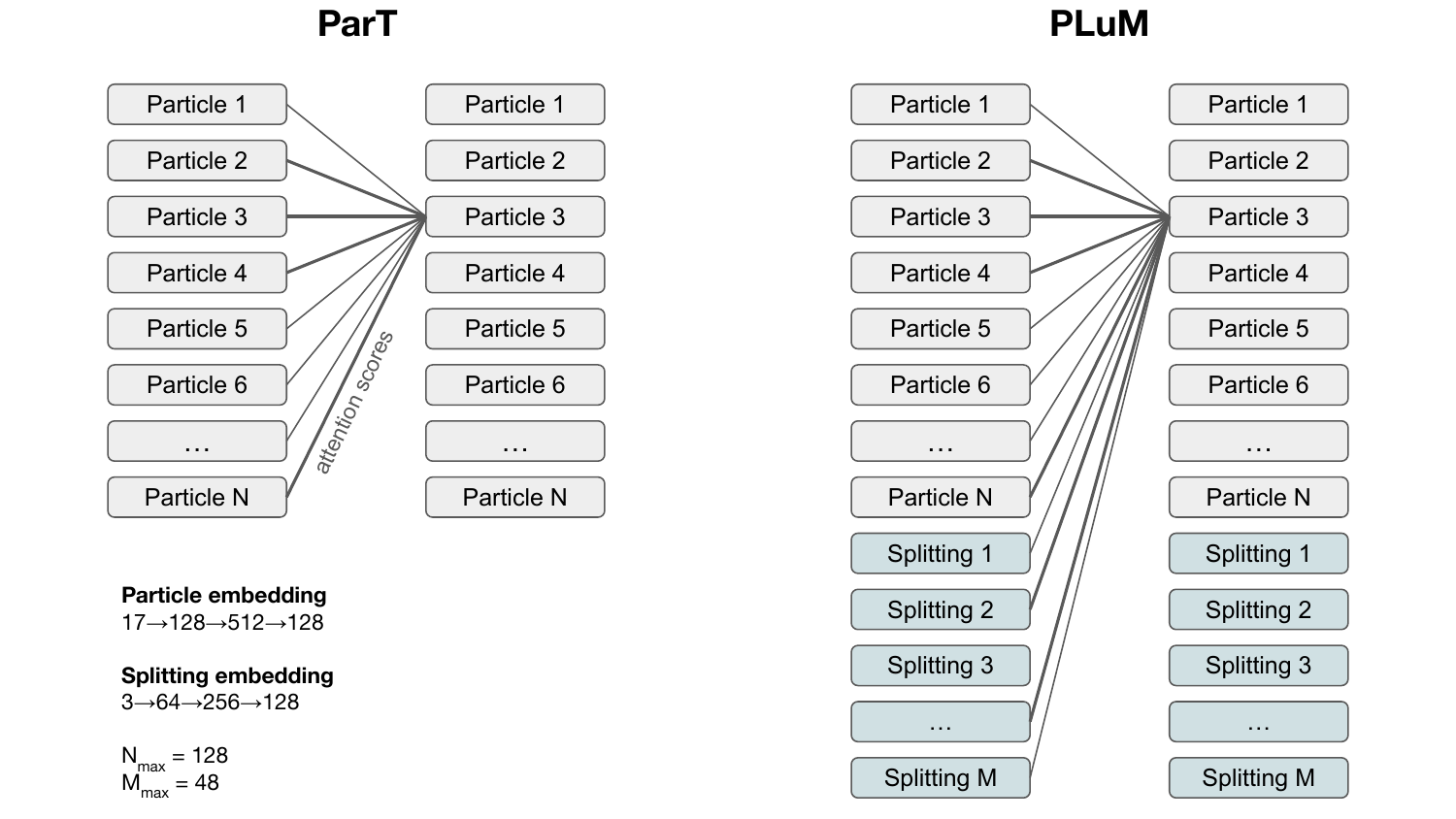}\\
\caption{Sketch of the attention mechanism in the encoder part of ParT (left) and, respectively, PLuM (right).}
\label{fig:sketch}
\end{centering}
\end{figure*}

Although both the particle features and the Lund plane splittings originate from the same jet constituents, they encode fundamentally different inductive biases and therefore provide distinct views of the same physical object. Particle-level inputs capture low-level kinematic information and pairwise relations directly, while the Lund representation explicitly organizes radiation into a hierarchical branching structure motivated by QCD dynamics. We therefore treat them as distinct input modalities, embedding them separately and concatenating their hidden representations right before the message passing step, i.e., the computation of attention scores across the full input sequence. This design follows the principles of multimodal learning, enabling the network to jointly reason over multiple structured views of the same physical object.

The Lund plane information, which is represented through the tuples 
\[
\mathcal{T}^{(i)} = \{\log k_\mathrm{T},\log \Delta R,\log z\}
\]
for each splitting $i$, is projected into the same latent space as the particle and pairwise embeddings via a lightweight multi-layer perceptron (MLP) with hidden-layer dimensions of $[64, 256, 128]$. Up to 48 splittings are considered per jet. The resulting embeddings are concatenated with the embedded particles and processed jointly by the transformer encoder. This integration allows the attention mechanism to correlate learned representations across constituent, pairwise, and Lund plane-derived features, without altering the underlying transformer structure itself, but implicitly extending the receptive field to include the augmented input space. In the ParT architecture, a bias term, derived from the pairwise features and representing the particle-particle interactions, is added to the attention matrix. When adding the splittings, we let the Lund tokens attend to all other elements in the sequence (other Lund tokens and the particles) and apply no prior bias. This way, the model learns to weigh Lund-plane-derived tokens purely based on learned self-attention dynamics. Overall, the change in the number of trainable parameters due to the separate embedding is modest, increasing from 2.14M for the default ParT configuration to 2.19M. We refer to this tagging algorithm as \emph{PLuM}, reflecting it incorporates \emph{P}article- and \emph{Lu}nd-views of the same jet in a \emph{M}ultimodal tagger. A sketch of the algorithm is presented in \autoref{fig:sketch}.

In the following, we focus on the $k_\mathrm{T}$-derived Lund plane. Of each network and configuration, ten copies are trained for 50 epochs in binary classification mode on the \textsc{JetClass}~\cite{Qu:2022mxj} dataset, a large open dataset of simulated jets reconstructed with the simplified detector simulation Delphes~\cite{deFavereau:2013fsa}. The trainings run over 16M (8M signal+8M background) jets per epoch, with a minibatch size of 256. Results are presented in terms of the maximum and mean performance of these trainings when evaluated on an independent test dataset of 2M+2M jets.


\section{Results}

\begin{figure*}[t!]
\begin{centering}
\includegraphics[width=0.475\textwidth]{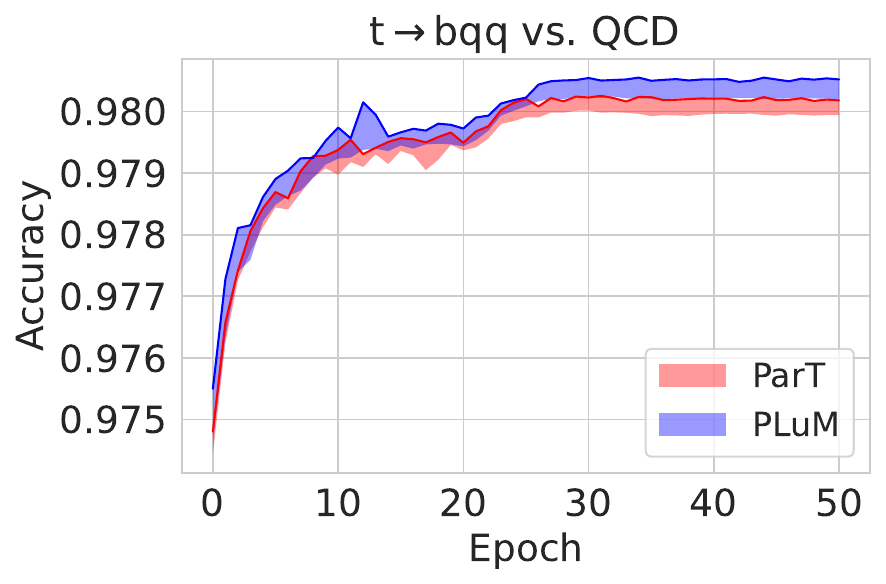}
\includegraphics[width=0.475\textwidth]{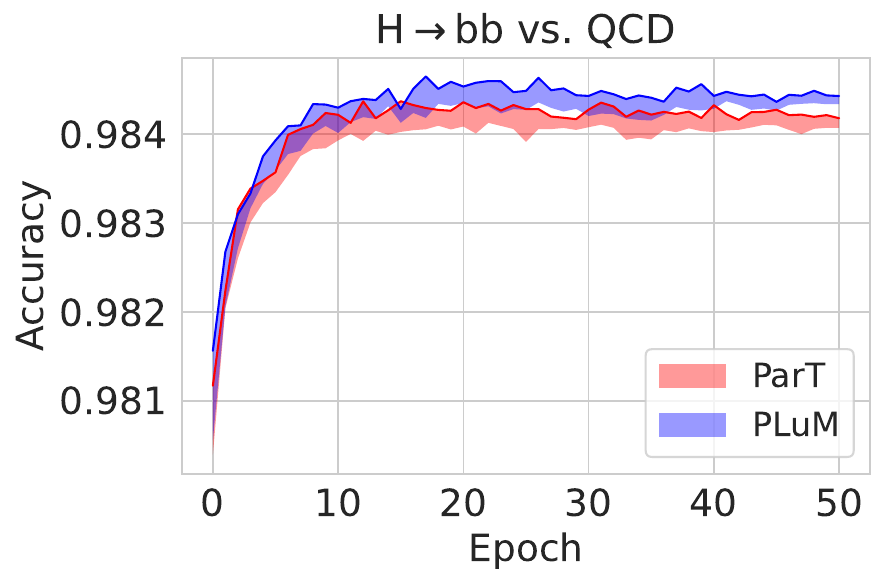}\\
\includegraphics[width=0.475\textwidth]{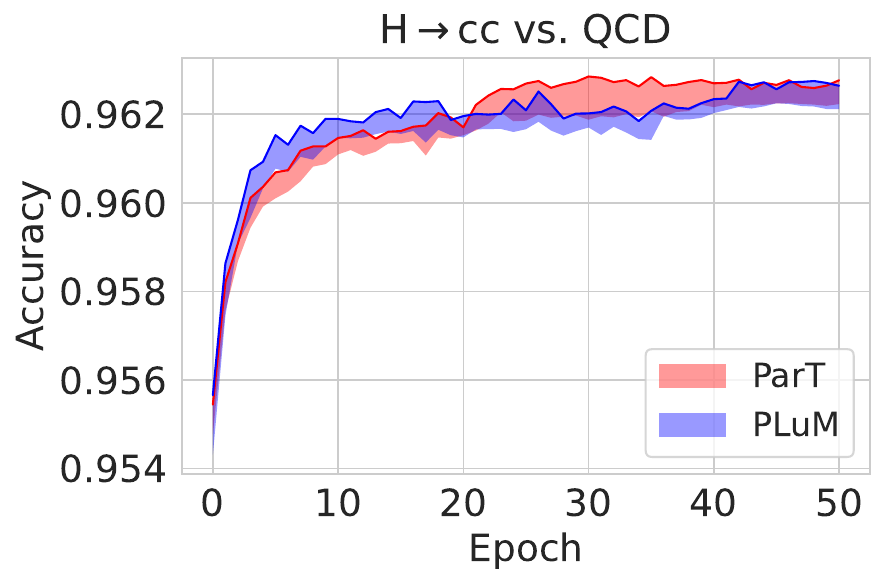}
\includegraphics[width=0.475\textwidth]{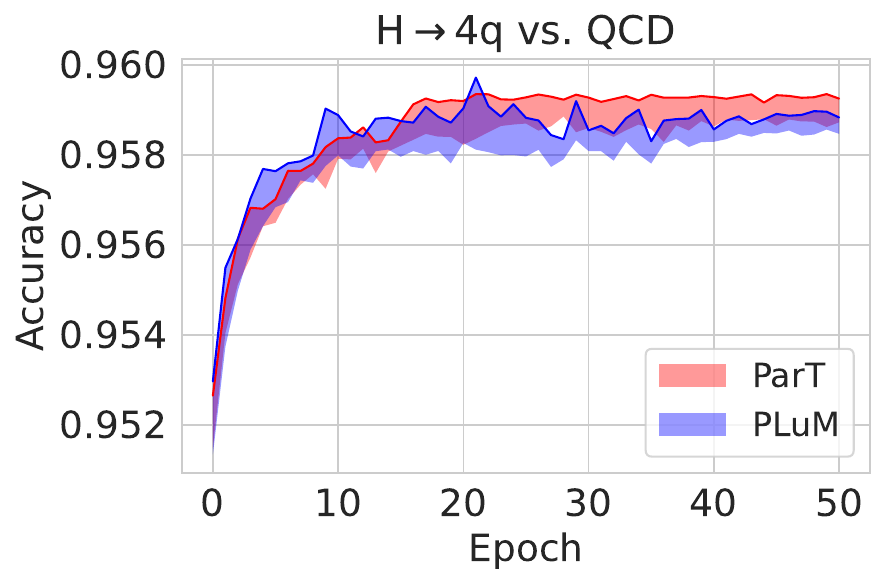}\\
\caption{Accuracies for the trainings of ParT (red) and PLuM (blue). Top left: top vs. QCD jets; top right: $\mathrm{H}\to\mathrm{b}\mathrm{\bar{b}}$ vs. QCD jets; bottom left: $\mathrm{H}\to\mathrm{c}\mathrm{\bar{c}}$ vs. QCD jets; bottom right: $\mathrm{H}\to\mathrm{WW^*}\to4\mathrm{q}$ vs. QCD jets. Solid lines indicate the maximum performance per epoch observed across 10 repeated trainings, while the bands indicate the difference w.r.t. the mean performance.}
\label{fig:acc}
\end{centering}
\end{figure*}

We evaluate the impact of incorporating physics-structured Lund plane splitting information into the ParT architecture across four distinct large-radius jet classification tasks. The performance trajectories over 50 training epochs are shown in \autoref{fig:acc}.

For both top-quark and $\mathrm{H}\rightarrow \mathrm{b}\bar{\mathrm{b}}$ tagging, the PLuM architecture demonstrates an immediate, systematic improvement in classification accuracy that remains stable throughout optimization. This robust gain persists well beyond the observed variation across repeated seeds, indicating a genuine enhancement in the learned latent representations rather than statistical fluctuations. As shown in \autoref{tab:wps}, we observe improved background rejection in $\mathrm{H}\rightarrow \mathrm{b}\bar{\mathrm{b}}$ tagging by up to 16\% at fixed signal efficiency despite only a marginal increase in model complexity. This means that a $(1.12)^2\sim$25\% higher background rejection could be achieved in boosted di-Higgs(4b) searches~\cite{CMS:2022gjd,ATLAS:2023qzf} for the same signal selection efficiency.

\renewcommand{\arraystretch}{1.3}
\begin{table}[thbp]
  \centering
  \caption{Average performance for the five best models for the $\mathrm{H}\to\mathrm{b}\bar{\mathrm{b}}$ vs. QCD and $\mathrm{t}\to\mathrm{bqq}$ vs. QCD classification task, respectively. The rejection power (Rej) is defined as the inverted false positive rate for a specific signal selection efficiency.}
  \begin{tabular}{@{}lcccc@{}}
    \toprule
    &\multicolumn{2}{c}{$\mathrm{H}\to\mathrm{b}\bar{\mathrm{b}}$ vs. QCD}&\multicolumn{2}{c}{$\mathrm{t}\to\mathrm{bqq}$ vs. QCD}\\
     & Rej\textsubscript{sig=50\%} & Rej\textsubscript{sig=90\%} &Rej\textsubscript{sig=50\%} & Rej\textsubscript{sig=90\%} \\
    \midrule
    ParT & 5864 & 386 & 13422 & 331\\
    PLuM & \textbf{6567} & \textbf{398} & \textbf{14388} & \textbf{353}\\
    \bottomrule
  \end{tabular}
  \label{tab:wps}
\end{table}
The observations imply that the sequential splittings natively captured by the $k_\mathrm{T}$-derived Lund plane tree explicitly encode the complex interplay of perturbative radiation, fragmentation, and subsequent displaced hadron decays characteristic of b-quark dynamics. By introducing this organizational structure as an explicit input modality, PLuM exploits the hierarchical radiation patterns alongside constituent kinematics, helping the model to isolate distinct heavy-flavor signatures.

Conversely, for $\mathrm{H}\rightarrow \mathrm{c}\bar{\mathrm{c}}$ and $\mathrm{H}\rightarrow 4$q topologies, the addition of Lund plane tokens yields no measurable benefit, exhibiting performance comparable to or slightly below the baseline architecture. 
This localized improvement suggests that the utility of physics-informed, hierarchical features is highly dependent on the physical properties of the underlying jet. 

The mechanisms underpinning this outcome differ between the two channels. For $\mathrm{H}\rightarrow \mathrm{c}\bar{\mathrm{c}}$ jets, the fragmentation and displaced heavy-hadron decays on the soft/wide-angle region of the Lund plane is substantially less pronounced than in $\mathrm{H}\rightarrow \mathrm{b}\bar{\mathrm{b}}$, as charm hadrons exhibit decay lengths roughly a factor of two to three shorter than those of B hadrons and produce correspondingly less distinctive soft radiation patterns. The PLuM training trajectory for this channel approaches the baseline only at late epochs, consistent with the model gradually learning to downweight Lund tokens that carry no complementary information beyond what particle-level attention already extracts. For $\mathrm{H}\rightarrow 4$q, the dominant effect is structural rather than dynamical: the four-prong $\mathrm{WW}^*$ topology populates the Lund tree more densely than two-prong decays, with about 17\% of signal jets exceeding the 48-splitting input cap, compared to about 9\% in $\mathrm{H}\rightarrow \mathrm{b}\bar{\mathrm{b}}$, presenting the encoder with a length-biased view of the radiation pattern. Combined with the absence of any heavy-flavor handle in the all-light-quark final state, the splitting modality contributes no compensating information, and the truncated tokens act as mild noise on the latent representation rather than as complementary structure.

\begin{figure*}[t!]
\begin{centering}
\includegraphics[width=0.975\textwidth]{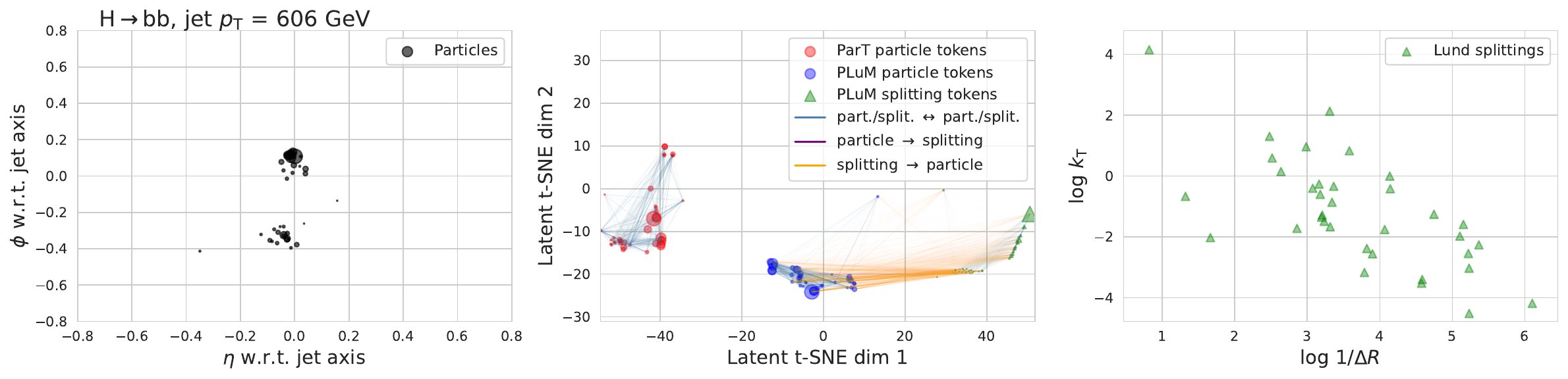}\\
\includegraphics[width=0.975\textwidth]{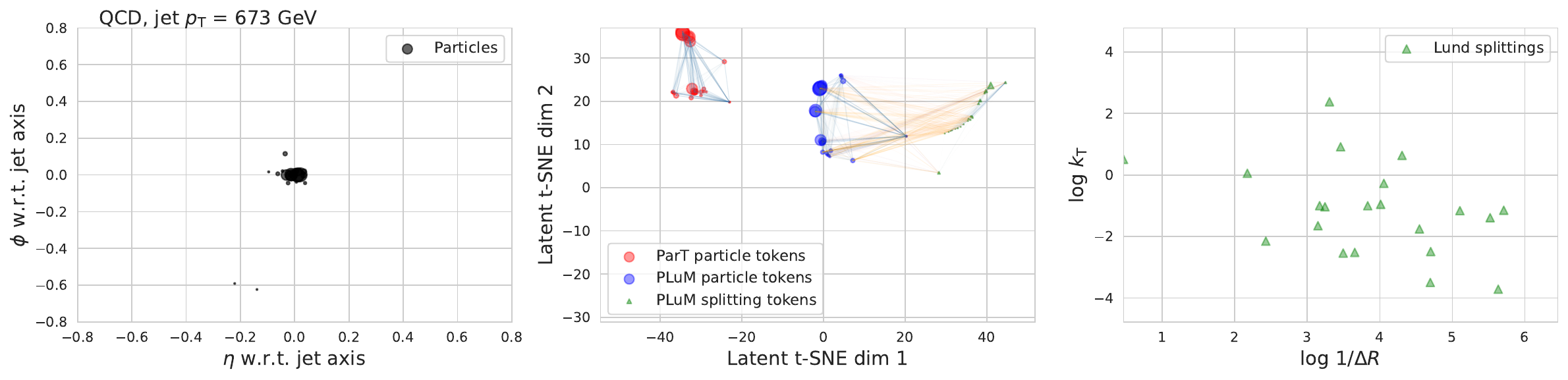}\\
\caption{Visualization of example $\mathrm{H}\to\mathrm{b}\bar{\mathrm{b}}$ (top) and QCD (bottom) jets in the detector space (left), the 128-dimensional latent space (middle), and, respectively, the Lund plane (right). For the left and middle column, the markers for the particle tokens are scaled according to their  $p_\mathrm{T}$, and the markers for the splitting tokens are scaled according to their $k_\mathrm{T}$. Attention scores are visualized as lines connecting the tokens. Blue lines correspond to the attention scores within the same family of tokens (particles or splittings, respectively), averaged over both directions. Purple and orange lines are the cross-attention scores between different modalities, in either direction. Attention scores are averaged over all eight attention heads.}
\label{fig:tsne}
\end{centering}
\end{figure*}


\autoref{fig:tsne} qualitatively explores the internal structure of the multimodal representations by visualizing an exemplary $\mathrm{H}\to\mathrm{b}\bar{\mathrm{b}}$ signal jet and a QCD background jet across the detector space, the model's latent space, and the Lund plane. In the detector space (left), final-state constituent particles are shown relative to the jet axis, with markers scaled by their transverse momentum ($p_{\mathrm{T}}$). The middle column presents a two-dimensional $t$-SNE projection~\cite{tsne} of the $128$-dimensional latent space after the encoder, contrasting the particle tokens of the baseline ParT architecture with the joint particle and splitting tokens of PLuM. 



The localized proximity of low-$k_{\mathrm{T}}$ splitting tokens to the particle tokens in the projected latent space of PLuM suggests that the model effectively correlates representations between soft/collinear radiation patterns and low-level constituent information. The attention patterns reveal that PLuM systematically assigns higher scores when the Lund splittings query the particles (orange lines in the middle plots of \autoref{fig:tsne}) than when particles query the splittings (purple). This asymmetry suggests the model learns to ground each QCD branching back onto the particles that generated it, while the small weights in the particle-to-Lund direction suggest that constituent particles do not strongly rely on explicit splitting information to form their representations, likely because the hierarchical branching structure is already implicitly encoded in the particle embeddings through self-attention. The observed gain therefore likely does not stem from particles being modified by splitting information, but rather by splitting tokens becoming more informative and robust features by attending to the relevant particles. This is then directly available to the classifier head, achieving better discrimination between signals with b quark content and background.

\begin{figure*}[t!]
\begin{centering}
\includegraphics[width=0.925\textwidth]{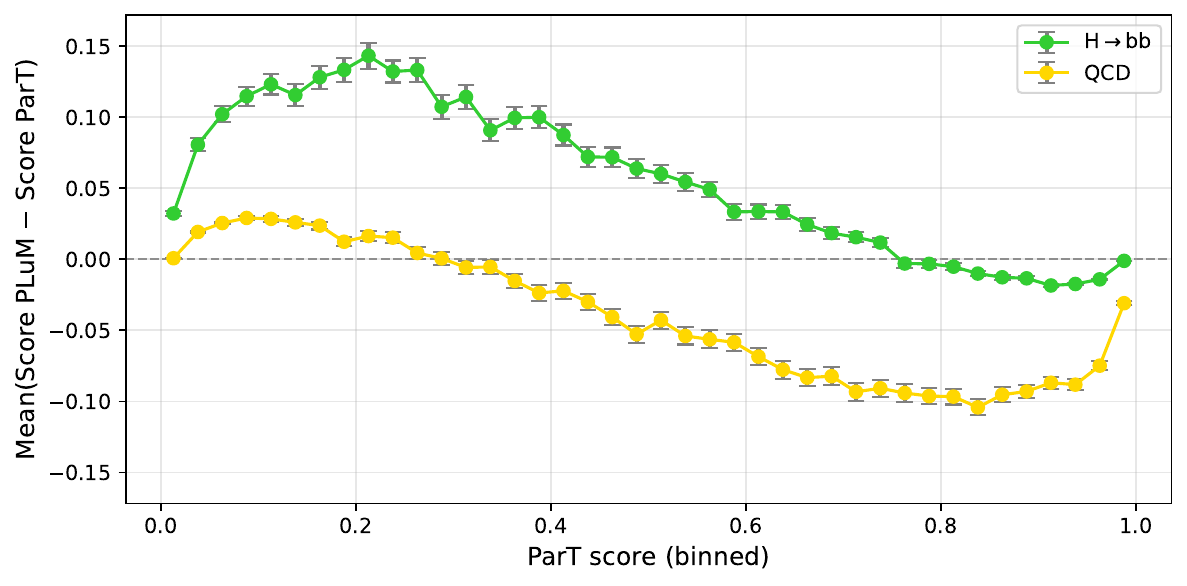}\\
\caption{Mean score difference for the same jets evaluated with the best PLuM and, respectively, best ParT models for the $\mathrm{H}\rightarrow \mathrm{b}\bar{\mathrm{b}}$ vs. QCD separation task, as a function of the ParT score.}
\label{fig:diff}
\end{centering}
\end{figure*}

~\autoref{fig:diff} indicates that the gain from PLuM does not arise from a uniform rescaling of classifier outputs, but from a systematic redistribution of uncertain events: especially for the cases with a ParT score in the range of 0.3--0.8, PLuM is consistently more confident for signal jets and shifts background jets toward lower scores. Near the extremes of the score distribution, the differences diminish, indicating that the additional hierarchical information mainly enhances decision boundaries rather than modifying already well-separated examples.


\section{Ablation studies and additional checks}

The following checks have been performed for the top quark discrimination task to better understand the observed gains:

\begin{itemize}
\item Using splittings from clustering with the CA algorithm did not result in a measurable performance increase compared to the ParT algorithm.
\item Increasing the number of attention heads from the ParT default 8 to 10 did not result in additional performance gains beyond what was observed for PLuM with 8 attention heads.
\item Including up to 96 splittings (instead of 48) did not result in extra performance gains.
\item Using the quintuple $\mathcal{T}^{(i)} = \{k_\mathrm{T},\Delta R,z,m,\psi\}$ instead of the triplet $\{k_\mathrm{T},\Delta R, z\}$ did not lead to a gain over the ParT algorithm.  
\end{itemize}

\section{Summary}
We investigated whether state-of-the-art particle transformers saturate the information content of particle-level jet representations or whether explicit representations of QCD radiation remain complementary. By augmenting particle transformers with Lund plane splittings, we observe systematic gains for heavy-flavor signatures while finding no comparable improvement for lighter topologies. These results suggest that hierarchical radiation information encoded in the Lund representation is not uniformly reconstructed from particle-level inputs alone, even in highly expressive transformer architectures. When applied to boosted Higgs tagging in the search for non-resonant $\mathrm{H}\mathrm{H}(4\mathrm{b})$ production, our algorithm leads to a 25\% larger background rejection than the current state of the art.
Future studies should assess whether the observed gains persist in experimental settings including realistic secondary-vertex information and detector effects. This was not checked because of the absence of reliable secondary vertex features in the fast simulation framework used in this study. Furthermore, the PLuM paradigm should be studied for narrow-cone jets, Lorentz-equivariant architectures, and searches for physics beyond the Standard Model. More generally, understanding in detail how transformer architectures utilize structured representations of QCD radiation may  provide broader insights into the design of future machine-learning methods for collider physics. 

\section*{Acknowledgements}
The trainings were performed on the Brown Oscar cluster, the MIT subMIT computing cluster, and the KIT ETP computing cluster. B. M. acknowledges the support from Schmidt Sciences. L. G. is supported by the DOE, Office of Science, Office of High Energy Physics Early Career Research program under Award No. DE-SC0026288.

\nocite{*}

\bibliography{apssamp}

\end{document}